\hsize=125mm
  \magnification=\magstep1
  \font\title = cmbx10 scaled \magstep2
  \overfullrule=0pt
5th February 1996    
\bigskip               
 
\bigskip

\bigskip

\bigskip

\bigskip

\bigskip

\bigskip

\bigskip

\bigskip

\bigskip

\centerline{\title Schr\"{o}dinger Equation of the Schwarzschild Black Hole}

\bigskip

\centerline{Jarmo M\"{a}kel\"{a}}

\medskip
 
\centerline{{\it D.A.M.T.P., Silver Street, Cambridge CB3 9EW, England}
\footnote*{e-mail: j.m.m.makela@damtp.cam.ac.uk}}

\bigskip

\centerline{\bf Abstract}

\medskip

\medskip

  We describe the gravitational degrees of freedom of the
Schwarzschild black hole by one free variable. We introduce an
equation which we suggest to be the Schr\"{o}dinger equation of the
Schwarzschild black hole corresponding to this model. We solve the
Schr\"{o}dinger equation explicitly and obtain the mass spectrum of the
black hole as such as it can be observed by an observer very far away
and at rest relative to the black hole.  Our equation implies that
there is no singularity inside the Schwarzschild black hole, and that the
black hole has a certain ground state in which its mass is non-zero. 

\centerline{}

\bigskip

\bigskip

\bigskip
     
    \centerline{\title 1. Introduction}

\medskip

\medskip

\medskip

          One of the basic requirements every physical theory must
satisfy is that the theory must be able to predict the possible
outcomes of measurements. In order to be a reliable physical theory
this requirement must be satisfied even by as esoteric a  theory as
quantum gravity.

          So far the quantum theory of gravity has given rather few
direct physical
predictions. Perhaps the most important of them are the existence of the so
called {\it Hawking radiation} emitted by black holes,$\lbrack 1\rbrack$ and
the result given
by Ashtekar, Rovelli and Smolin, which says that area is
quantized.$\lbrack 2\rbrack$ Quite a lot of effort has been spent on
the study of
quantum cosmology. An application of quantum gravity to quantum
cosmology, however, meets with grave conceptual difficulties, such as
the interpretation of the wave function if the observers are assumed
to be a part of the physical system under consideration, and the problem of
time.$\lbrack 3\rbrack$

         In this paper an attempt is made to use quantum gravity in
order to predict the physical properties of the Schwarzschild black hole
as such as they are observed by an observer who is very far away 
and at rest relative to the black hole. We assume that the
Schwarzschild black hole is a quantum mechanical system, and that its
physical states can therefore be described by a certain wave function
$\psi$. Since we assume that the observer is very far away from the
black hole, he can consider himself, at least to a very good
approximation, as an observer external to the physical system, the
black hole, under consideration. Moreover, spacetime around him is, at
least to a very good approximation, flat, and so he can use the
standard interpretation of quantum mechanics, and he has a
well-defined time coordinate. In other words, the conceptual problems
related to quantum gravity are absent.

        At least classically, the only thing an external observer can
observe on the properties of the Schwarzschild black hole is its mass
$M$. Because of that, our object is to predict the possible masses of
the Schwarzschild black hole as such as they are observed by an
external observer very far away and at rest relative to the black
hole. The mass $M$ measured by this kind of an observer is the same mass
as the one written in the usual expression of the Schwarzschild
metric. In terms of the mass $M$, the
observer can define the concept of energy $E$ of the black hole as:
$$
E=Mc^2,\eqno(1.1)
$$  
where $c$ is the velocity of light. It is obvious that if the observer
can predict the energies of the black hole, he can also predict its
possible masses.

        In order to predict the energies of the black hole, the
observer writes down the time-independent Schr\"{o}dinger equation
$$
\hat{H}\psi = E\psi\eqno(1.2)
$$
of the black hole. In this equation, $\hat{H}$ is the Hamiltonian
operator of the black hole, and $\psi$ is its wave function. The
possible outcomes $E_n$ of the measurements of the energy are the
eigenvalues of $\hat{H}$, and the eigenfunctions $\psi_n$ give the
corresponding propability amplitudes. In other words, if we can
construct the Schr\"{o}dinger equation of the Schwarzschild black hole, we
can predict its possible masses.

  In this paper we describe the gravitational degrees of freedom
of the black hole by one free variable. We introduce an equation which
we suggest to be the time-independent
Schr\"{o}dinger equation of the Schwarzschild black hole corresponding to
our model as such as it can
be used by an observer very far away and at rest relative to the black hole.
It turns out that our Schr\"{o}dinger equation can be solved explicitly,
and so we can predict the mass spectrum of the Schwarzschild black
hole. The mass spectrum, in turn, can be calculated if we can calculate
the spectrum of the area 
$$
A_S := {{16\pi G^2}\over{c^4}}M^2,\eqno(1.3)
$$
of the event horizon of the Schwarzschild black hole. It turns out
that the eigenvalues of $A_S$ are an integer plus $(1\slash 2)$ times a
certain area, whose order of magnitude is $10^{-68}m^2$. This result
is entirely in harmony with the results of Ashtekar, Rovelli and Smolin
on the quantization of area in quantum gravity.$\lbrack 2\rbrack$ It is
also compatible
with the studies made by Bekenstein and others on the properties of
black holes.$\lbrack 4-8\rbrack$

     We shall also see that our Schr\"{o}dinger equation, if true, solves the
fundamental problem on whether there is a singularity inside the
Schwarzschild black hole. The answer turns out to be negative, since
if there were a singularity inside the black hole, then our
Schr\"{o}dinger equation would have no physically acceptable solutions.
Moreover, it turns out that
there is a certain ground state in which the mass of the black hole is
non-zero.

\bigskip

\bigskip

\bigskip

 \centerline{\title 2. Schr\"{o}dinger Equation}

\bigskip

\bigskip

          When searching for the Schr\"{o}dinger equation of the
Schwarzschild black hole, one must first find its classical
Hamiltonian from the point of view of an observer at rest very far
away. This problem brings us to the Hamiltonian dynamics of
asymptotically flat spacetimes.

       An extensive study of the Hamiltonian dynamis of asymptotically
flat spacetimes was made long ago by Regge and Teitelboim[9]. They found
that in asymptotically flat spacetimes certain {\it surface integrals}
at spatial infinity play a decisive role. For example, the true Hamiltonian  of
an asymptotically flat spacetime is not the Hamiltonian $H_0$ of
spatially compact spacetimes written in terms of the lapse $N$, the
shift $N^i$, and the Hamiltonian and the diffeomorphism constraints
${\cal H}$ and ${\cal H}_i$ as:
$$
H_0 := \int d^3x(N{\cal H} + N^i{\cal H}_i),\eqno(2.1)
$$
but the correct Hamiltonian is
$$
H := H_0 + E_{ADM},\eqno(2.2)
$$
where $E_{ADM}$ is the so called {\it ADM energy}. If the spacetime
coordinates have been chosen in such a way that the spacetime metric
$g_{\mu\nu}$ becomes to the flat Minkowski metric $\eta_{\mu\nu}$ at
spatial infinity, then $E_{ADM}$ can be written as a surface integral
at spatial infinity:
$$
E_{ADM} = {{c^4}\over{16\pi G}} \oint d^2s_k(g_{ik,i} -
g_{ii,k}),\eqno(2.3)
$$
where $i,k=1,2,3$. It can be easily shown that for the Schwarzschild
black hole at rest we have:
$$
E_{ADM} = Mc^2.\eqno(2.4)
$$
Indeed, this can be considered as the energy of the black hole. In
this paper, when we talk about the energy of the Schwarzschild black
hole, we always mean its ADM energy.

    It was one of the main results of Regge and Teitelboim that if one
fixes the coordinate system by fixing the lapse $N$ and the shift
$N^i$ then, in this fixed coordinate system, the correct Hamiltonian of
an asymptotically flat spacetime is obtained by inserting the solution
of the Hamiltonian and the diffeomorphism constraints
$$
\eqalignno{ {\cal H} &=0,&(2.5.a)\cr
           {\cal H}_i&=0,&(2.5.b)\cr}
$$
into the surface integral (2.3). More precisely, between the phase
space coordinates of spacetime and the numerical value of the ADM energy
(2.3) there is a certain relationship which can be solved from the
constraints (2.5) in the fixed coordinate system. If the numerical
value of the ADM energy is written in terms of the phase space
coordinates, by using the relationship found from the constraints
(2.5), we get the spacetime Hamiltonian in our coordinate system. In
our case, this means that we must first find an appropriate variable
$a$ describing the gravitational degrees of freedom of the
Schwarzschild black hole, and then write the constraints (2.5) in
terms of $a$ and its canonical momentum $p$. From the constraints
(2.5) we find how the quantity $Mc^2$, which is the ADM energy of our 
spacetime, depends on $a$ and $p$. In this way we get the classical 
Hamiltonian of the Schwarzschild black hole.  

       To begin with, we write down the spacetime metric of the
Schwarzschild black hole in terms of the Schwarzschild coordinates
$r$ and $t$, and the spherical angles $\theta$ and $\phi$. Outside the
black hole horizon the metric is:
$$
ds^2 = -\left(1- {{2GM}\over{c^2 r}}\right)c^2\,dt^2 +
{1\over{1-{{2GM}\over{c^2r}}}}\,dr^2 + r^2(d\theta^2 +
\sin^2\theta\,d\phi^2),\eqno(2.6.a)
$$
whereas inside the black hole horizon the metric is:
$$
ds^2 = \left({{2GM}\over{c^2r}} - 1\right)c^2\,dt^2 -
{1\over{{{2GM}\over{c^2r}} - 1}}\,dr^2 + r^2(d\theta^2 +
\sin^2\theta\,d\phi^2). \eqno(2.6.b)
$$
We are outside the black hole horizon when $r$ is greater than the
Schwarzschild radius
$$
R_S := {{2GM}\over{c^2}},\eqno(2.7)
$$
and we are inside the black hole horizon when $r$ is smaller than the
Schwarzschild radius.

    We must now find an appropriate coordinate system. Since the
coordinates $r$ and $t$ behave remarkably badly when $r=R_S$, and
inside the black hole horizon $t$ is not a timelike coordinate, the
coordinates $r$ and $t$ are out of question. We have plenty of choice
because the ADM energy is independent of the chosen coordinate system
provided that the coordinate system is at rest at spatial infinity
relative to the black hole, and the time coordinate coincides with the
time coordinate $t$ at infinity. In this regard, an appropriate
choice is to use the so called {\it Novikov coordinates}[10], and we
begin with a brief review on the properties of these coordinates.

    The basic idea of the Novikov coordinates is to relate the
spacetime coordinates to observers in a radial free fall towards the
center of the black hole such that all these observers are at rest
relative to the black hole when $t=0$, and at that moment of the time
$t$ they are released into a free fall. The proper times of these
observers give the time coordinate, and a certain variable related to
the positions of these obsevers when $t=0$, gives one of the spatial
coordinates for each point in space and time. Since an observer in a
radial free fall is always at rest relative to the black hole at
spatial infinity, and his proper time coincides with the time $t$ at
spatial infinity, the Novikov coordinates behave in the desired way.

       It can be easily shown that for an observer in a radial free fall
we have, in general:
$$
\left(1-{{2GM}\over{c^2 r}}\right)\dot{t} = constant:= {\chi\over{\sqrt{1+\chi^2}}}, \eqno(2.8)
$$     
where the dot means proper time derivative, and the constant $\chi\ge
0$. Because of that, we find that when $r$ goes from $r$
to $r-dr$, then the proper time $\tau$ of the observer in a free fall
goes from $\tau$ to $\tau+d\tau$ such that
$$
c^2\,d\tau^2 = - {{{\chi^2}\over{1+\chi^2}}\over{{{2GM}\over{c^2 r}} - 1}}c^2\,d\tau^2 +
{{dr^2}\over{{{2GM}\over{c^2 r}} - 1}},\eqno(2.9)
$$
and we find that the equation of motion of the observer in a free fall
is
$$
\dot{r}^2 = {{2GM}\over r} - {{c^2}\over{1+\chi^2}}.\eqno(2.10)
$$
As one can see, the observer is at rest relative to the black hole at
the point where 
$$
r=r_{max}:= (1+\chi^2){{2GM}\over{c^2}},\eqno(2.11)
$$
and so the constant $\chi$ can be written in terms of $r_{max}$ as:
$$
\chi = \left( {{r_{max}}\over{R_S}} - 1\right)^{1\slash 2}.\eqno(2.12)
$$

     The Novikov coordinates are now the coordinates $\tau$ and $\chi$
such that $r=r_{max}$ and $t=0$ when $\tau=0$. Using Eqs.(2.8) and
(2.10) we can, at least in principle, express the coordinates $r$ and
$t$ in terms of $\tau$ and $\chi$. In other words, we have
$r=r(\tau,\chi)$, and $t=t(\tau,\chi)$. Because we have:
$$      
\eqalignno{dt &={{\partial t}\over{\partial \tau}}\,d\tau + {{\partial
t}\over{\partial\chi}}\,d\chi,&(2.13.a)\cr
           dr &= {{\partial r}\over{\partial\tau}}\,d\tau + {{\partial
r}\over{\partial\chi}}\,d\chi,&(2.13.b),\cr}
$$
we find from Eq.(2.6) that if we can express $r$ and $t$ in terms of
$\tau$ and $\chi$, the spacetime metric becomes to:
$$
\eqalign{ds^2 = -\left(1-{{2GM}\over{c^2 r}}\right)\left(c{{\partial
t}\over{\partial \tau}}\,d\tau + c{{\partial
t}\over{\partial\chi}}\,d\chi\right)^2 &+
{1\over{1-{{2GM}\over{c^2r}}}}\left( {{\partial r}\over{\partial
\tau}}\,d\tau + {{\partial r}\over{\partial\chi}}\,d\chi\right)^2\cr 
                                       &+r^2(d\theta^2 + 
\sin^2\theta\,d\phi^2).\cr}\eqno(2.14)
$$
The precise relationship between $r$, $\tau$ and $\chi$, although in
an implicit form, can be calculated from Eq.(2.10). We get:
$$
\tau={1\over c}(1+\chi^2)\left(R_S r -
{{r^2}\over{1+\chi^2}}\right)^{1\slash 2} + {{R_S}\over
c}(1+\chi^2)^{3\slash 2} \cos^{-1}\left[\left({{r\slash
R_S}\over{1+\chi^2}}\right)^{1\slash2}\right].\eqno(2.15)
$$
Differentiating the both sides of this equation with respect to $\chi$
we get an expression for $(\partial r\slash\partial\chi)$ in terms of
$r$ and $\chi$:
$$
{{\partial r}\over{\partial\chi}} = 3R_S\chi - {{\chi
r}\over{1+\chi^2}} + 3R_S\chi(1+\chi^2)^{1\slash 2}\left({{R_S}\over r} -
{1\over{1+\chi^2}}\right)^{1\slash 2}\cos^{-1}\left[\left({{r\slash
R_S}\over{1+\chi^2}}\right)^{1\slash 2}\right],\eqno(2.16)
$$
and from Eqs.(2.8) and (2.10) we get:
$$
\eqalignno{{{\partial t}\over{\partial\tau}} &= {1\over{1-{{R_S}\over
r}}}{\chi\over{\sqrt{1+\chi^2}}},&(2.17.a)\cr
           {{\partial r}\over{\partial\tau}} &= -c\left({{R_S}\over r}
- {1\over{1 + \chi^2}}\right)^{1\slash 2}.&(2.17.b)\cr}
$$

   The quantity $(\partial t\slash\partial\chi)$ can also be expressed
in terms of $r$ and $\chi$. As it is well known, the relationship
between $t$ and $r$ can be expressed in a parametrized form[10]:
$$
\eqalignno{r&={1\over 2}R_S(1+\chi^2)(1+\cos\eta),&(2.18.a)\cr
         t&={{R_S}\over c}\ln\left\vert{{\chi + \tan({\eta\over
2})}\over{\chi - \tan({\eta\over 2})}}\right\vert + {{R_S}\over
c}\chi[\eta + {1\over 2}(1+\chi^2)(\eta + \sin\eta)].&(2.18.b)\cr}
$$ 
If one solves the parameter $\eta$ from Eq.(2.18.a) in terms of $r$
and $\chi$, and inserts the result into Eq.(2.18.b), one gets an
expression for $t$ in terms of $r$
and $\chi$. From that expression one can calculate 
$(\partial t\slash\partial\chi)$ in terms of $\chi$, $r$ and 
$(\partial r\slash\partial\chi)$, and with the help of Eq.(2.16) one gets
$(\partial t\slash\partial\chi)$ in terms of $r$ and $\chi$.
If one inserts Eqs.(2.16) and (2.17), and the expression of $(\partial
t\slash\partial\chi)$ into the
metric (2.14), one gets an expression of the metric
written in terms of $r$ and $\chi$. It turns out to be[10]:
$$
ds^2 = -c^2\,d\tau^2 + {{1+\chi^2}\over{\chi^2}}\left({{\partial
r}\over{\partial \chi}}\right)^2\,d\chi^2 + r^2(d\theta^2 +
\sin^2\theta\,d\phi^2),\eqno(2.19) 
$$
and from Eq.(2.16) it follows that the metric takes the form:
$$
\eqalign{&ds^2 = -c^2\,d\tau^2 + (1+\chi^2)\cr
         &\times\left\lbrace 3R_S 
- {r\over{1+\chi^2}} + 3R_S(1+\chi^2)^{1\slash 2}\left({{R_S}\over r} -
{1\over{1+\chi^2}}\right)^{1\slash 2}\cos^{-1}\left[\left({{r\slash
R_S}\over{1+\chi^2}}\right)^{1\slash 2}\right]\right\rbrace^2\,d\chi^2\cr
&+ r^2(d\theta^2 + \sin^2\theta\,d\phi^2).\cr}\eqno(2.20)
$$
As one can see, this metric behaves perfectly well in the black hole
horizon $r=R_S$. When written in this form, it does not involve any
explicit $\tau$-dependence, but all $\tau$-dependence is included into
the function $r(\tau,\chi)$. 
      
   We have now found an appropriate coordinate system for the
whole spacetime. Our next task is to find an appropriate variable
describing the geometrical degrees of freedom of the Schwarzschild
black hole, and to write the spacetime metric in terms of this
variable. This variable should contain all $\tau$-dependence of the
metric. 

     We define the variable $a(\tau)$ describing the geometrical
degrees of freedom of the Schwarzschild black hole as:
$$
a(\tau) := r(\tau, 0),\eqno(2.21)
$$
for all $\tau\ge 0$. This definition can be understood as a boundary
condition to the differential equation (2.16). If one solves Eq.(2.16)
with the boundary condition (2.21), one gets $r(\tau,\chi)$ in
terms of $a(\tau)$ and $\chi$. It is easy to see that the general
solution of Eq.(2.16) is:
$$
(1+\chi^2)\left(R_S r -
{{r^2}\over{1+\chi^2}}\right)^{1\slash 2} + R_S(1+\chi^2)^{3\slash 2} \cos^{-1}\left[\left({{r\slash
R_S}\over{1+\chi^2}}\right)^{1\slash2}\right] = constant := 
{\cal C},\eqno(2.22)
$$
and therefore the solution satisfying the boundary condition (2.21)
is:
$$
\eqalign{&(1+\chi^2)\left(R_S r -
{{r^2}\over{1+\chi^2}}\right)^{1\slash 2} + R_S(1+\chi^2)^{3\slash 2} \cos^{-1}\left[\left({{r\slash
R_S}\over{1+\chi^2}}\right)^{1\slash2}\right]\cr 
&= (R_Sa - a^2)^{1\slash 2} 
+ R_S\cos^{-1}\left[\left({a\over{R_S}}\right)^{1\slash 2}\right].\cr}
\eqno(2.23)
$$

   To gain some insight into the geometric meaning of the variable
$a(\tau)$, let us write the metric when $\chi=0$. Since it
follows from Eqs.(2.16) and (2.17) that 
$$
\eqalignno{{{\partial r}\over{\partial\chi}} &= {{\partial
t}\over{\partial\tau}} = 0,&(2.24.a)\cr
           {{\partial r}\over{\partial\tau}} &= -c\left({{R_S}\over r}
- 1\right)^{1\slash 2},&(2.24.b)}
$$
when $\chi=0$, it follows from Eq.(2.14) and from the boundary condition
(2.21) that the metric can be written as:
$$
ds^2 = -c^2\,d\tau^2 + \left({{2GM}\over{c^2a(\tau)}} -
1\right)\left({{\partial t}\over{\partial\chi}}\right)^2c^2\,d\chi^2 +
a^2(\tau)(d\theta^2 + \sin^2\theta\,d\phi^2).\eqno(2.25)
$$
If we assume that $0<a(\tau)<R_S$, we can use $t$ as one of the
spatial coordinates, and we get:
$$         
ds^2 =-c^2\,d\tau^2 + \left({{2GM}\over{c^2a(\tau)}} -
1\right)c^2\,dt^2 +
a^2(\tau)(d\theta^2 + \sin^2\theta\,d\phi^2).\eqno(2.26)
$$
Comparing Eqs.(2.6.b) and (2.26) we find that the quantity $a(\tau)$
is, actually, the radius of the throat of the black hole as such as it
is observed by an observer inside the black hole horizon. We use
$a(\tau)$ as a minisuperspace-type variable of our model, and the idea
is that $a(\tau)$ carries, through Eq.(2.23), all information about the
time evolution of the spacelike hypersurface $\tau=constant$ of
spacetime. In other words, if we know how $a$ depends on the chosen
time coordinate $\tau$, we can calculate from Eq.(2.23) the value of
the function $r(\tau,\chi)$ for every $\tau$ and $\chi$. Inserting
$r(\tau,\chi)$ into the metric (2.20) we obtain the time evolution of
the spacelike hypersurface. We assume that $a$ can be any function of
the time coordinate $\tau$. However, if Eistein's equations written in 
terms of $a(\tau)$
are satisfied, then the precise relationship between $a(\tau)$ and
$\tau$ is the same as the one between $r(\tau, 0)$ and $\tau$ in
Eq.(2.15). Since it follows from Eq.(2.11) that, classically, $a(\tau)$ is
always smaller than, or equal to, the Schwarzschild radius $R_S$ of the
black hole, we find that $a(\tau)$ is an ideal choice for a variable
describing the gravitational degrees of freedom inside the black hole horizon. 

      Our next task is to write the constraints (2.5) and to find an
expression of the ADM energy $Mc^2$ in terms of $a$ and its canonical
momentum $p$. The Hamiltonian
constraint of spacetime without matter fields is, in general:
$$
{\cal H} = {{c^4}\over{16\pi G}}\sqrt{q}(K_{ij}K^{ij} - K^2 - {\cal
R}) 
= 0,\eqno(2.27)
$$
where $q$ is the determinant of the metric on the hypersurface
$\tau=constant$, $K_{ij}$ is the exterior curvature tensor, $K$ its
trace, and ${\cal R}$ is the Riemannian scalar on the hypersurface. We
find, by using the metric (2.19), that ${\cal R}$ and
$K_{ij}K^{ij}-K^2$ can be written as:
$$
\eqalignno{{\cal R} &= -{4\over
r}{{\chi}\over{(1+\chi^2)^2}}\left({{\partial r}\over{\partial
\chi}}\right)^{-1} + {2\over{r^2}}{1\over{1+\chi^2}},&(2.28.a)\cr
 K_{ij}K^{ij} - K^2 &= -{4\over{c^2r}}\left({{\partial r}\over{\partial
\chi}}\right)^{-1}{{\partial
r}\over{\partial\tau}}{\partial\over{\partial \tau}}\left({{\partial
r}\over{\partial\chi}}\right) - {2\over{c^2r^2}}\left({{\partial
r}\over{\partial\tau}}\right)^2.&(2.28.b)\cr}
$$
Because all time dependence is included in $a$, we have:
$$
{{\partial r}\over{\partial\tau}} = {{\partial r}\over{\partial
a}}\dot{a},\eqno(2.29)
$$
and because it follows from Eq.(2.23) that:
$$
{{\partial r}\over{\partial a}} = \left({{{{R_S}\over r} -
{1\over{1+\chi^2}}}\over{{{R_S}\over a} - 1}}\right)^{1\slash
2},\eqno(2.30)
$$
we find that the Hamiltonian constraint can be written in an arbitrary
point on the spacelike hypersurface $\tau=constant$ as:
$$
\eqalign{&{{3c^4}\over{8\pi G}}\sin\theta\left\lbrace
{{R_S}\over{(1+\chi^2)^{1\slash 2}}} - {r\over{(1+\chi^2)^{3\slash
2}}} + R_S\left({{R_S}\over r} - {1\over{1+\chi^2}}\right)^{1\slash
2}\cos^{-1}\left[\left({{r\slash R_S}\over{1+\chi^2}}\right)^{1\slash
2}\right]\right\rbrace\cr
&\times \left[{1\over{c^2}}\left({{R_S}\over a} -
1\right)^{-1}\dot{a}^2 - 1\right] = 0.\cr}\eqno(2.31)
$$
The solution of this constraint is, in every hypersurface point:
$$
\dot{a}^2 = c^2\left({{R_S}\over a} - 1\right),\eqno(2.32)
$$
and we find that the mass $M$ of the black hole can be written in
terms of $a$ and its time derivative $\dot{a}$ as:
$$
M={1\over{2G}}(a\dot{a}^2 + c^2a).\eqno(2.33)
$$
The diffeomorphism constraints do not give us anything new, and we find
that the ADM energy, and hence the classical Hamiltonian $H$ of the
Schwarzschild black hole, can be written in terms of $a$ and $\dot{a}$
as:
$$
H = {{c^2}\over{2G}} a\dot{a}^2 + {{c^4}\over{2G}}a.\eqno(2.34)
$$
The first term on the right hand side can now be considered as the
``kinetic energy'', and the second term as the ``potential energy'' of
the black hole, and therefore the black hole Lagrangian is:
$$
L = {{c^2}\over{2G}} a\dot{a}^2 - {{c^4}\over{2G}}a.\eqno(2.35)
$$
The canonical momentum conjugate to $a$ is therefore
$$
p := {{\partial L}\over{\partial\dot{a}}} = {{c^2}\over{G}}a\dot{a},\eqno(2.36)
$$ 
and the classical Hamiltonian of the Schwarzschild black hole takes in
terms of $a$ and $p$ a form:
$$
H = {G\over{2c^2}}{1\over a} p^2 + {{c^4}\over{2G}}a.\eqno(2.37)
$$ 

    We can now obtain the time-independent Schr\"{o}dinger equation
(1.2) of the
Schwarzschild black hole by replacing the classical Hamiltonian $H$
by the corresponding operator $\hat{H}$. To find a correct expression
to the operator $\hat{H}$ we must first specify the inner product
between the states $\vert\psi\rangle$ represented by the wave
functions $\psi=\psi(a)$ of the black hole. 
Since $a$ can be
thought to describe the distance from the center of the black hole,
a natural inner product between arbitrary states $\psi_1$ and $\psi_2$
is:
$$
\langle\psi_1\vert\psi_2\rangle := \int_0^\infty
\psi^{*}_1(a)\psi_2(a)a^2\,da.\eqno(2.38)
$$
As it is well known from elementary quantum mechanics, the correct
expression to the operator $\hat{p}^2$ corresponding to this kind of
inner product is:$\lbrack 11\rbrack$
$$
\hat{p}^2 = -\hbar^2\left({{d^2}\over{da^2}} + {2\over
a}{d\over{da}}\right),\eqno(2.39)
$$
and therefore we get:
$$   
\left\lbrack -{{\hbar^2G}\over{2c^2}} {1\over
a}\left({{d^2}\over{da^2}} + {2\over a}{d\over{da}}\right) +
{{c^4}\over{2G}} a\right\rbrack\psi(a) = E\psi(a).\eqno(2.40)
$$
We suggest that, within our minisuperspace-type model, this is the 
time-independent Schr\"{o}dinger equation of the Schwarzschild black hole.
  
\bigskip

\bigskip

\bigskip

\centerline{\title 3. Eigenvalues and Eigenstates}

\medskip

\medskip

\medskip

      Our next task is to solve the time-independent Schr\"{o}dinger
equation (2.40) of the
Schwarzschild black hole. To begin with, we
denote:
$$
\psi(a) := {1\over a}u(a),\eqno(3.1)
$$
and we get:
$$
\left\lbrack -{1\over{2k^4}}{{d^2}\over{da^2}} + {1\over 2}(a^2 -
R_Sa)\right\rbrack u(a) = 0,\eqno(3.2)
$$
where $k$ is, essentially, the inverse of the Planck length:
$$
k := \sqrt{{c^3}\over{\hbar G}},\eqno(3.3)
$$
and $R_S$ is the Schwarzschild radius. If we denote:
$$
\xi := a - {1\over 2}R_S,\eqno(3.4)
$$
we get a very interesting equation:
$$
\left(-{1\over{2k^4}}{{d^2}\over{d\xi^2}} + {1\over
2}\xi^2\right)u(\xi) = {1\over 8}R_S^2u(\xi).\eqno(3.5)
$$
As one can see, we have obtained the eigenvalue equation to the quantity
$(1\slash 8)R_S^2$, and hence, in essence, to the area $A_S$ of the
event horizon of the Schwarzschild black hole. Moreover, this equation
is something every physicist knows very well. It is the Schr\"{o}dinger
equation of linear harmonic oscillator. Its solutions are, in general,
of the form:
$$
u_n(\xi) = N_nH_n(k\xi)e^{-{1\over 2}k^2\xi^2},\eqno(3.6)
$$
where $n=0,1,2...$, $H_n$'s are Hermite polynomials, and $N_n$ is
an appropriate normalization factor. The corresponding eigenvalues of
$(1\slash 8)R_S^2$ are:
$$
{1\over 8}R_S^2(n) := (n+{1\over 2}){1\over{k^2}}.\eqno(3.7)
$$
In other words, the area of the event horizon of the Schwarzschild
black hole has a discrete spectrum.

     At this point, however, we meet with a very delicate problem,
which is related to the boundary conditions of the function $u(a)$. It
follows from Eq.(3.1) defining the function $u(a)$ that when $a$
goes to zero, then $u(a)$ must go to zero at least as fast as $a$,
since otherwise the wave function $\psi(a)$ would have a singularity
when $a=0$. When $u$ is written in terms of $\xi$, it follows from
Eqs.(3.4) and (3.7) that in the state $n$ the function $u_n(\xi)$
must go to zero in the point
$$
\xi_n:=-{1\over 2}R_S(n)=-\sqrt{2n+1}{1\over k}\eqno(3.8)
$$ 
at least as fast as the function $\xi+(1\slash 2)R_S(n)$. However,
none of the solutions $u_n(\xi)$ expressed in  Eq.(3.6) to the
eigenvalue equation
(3.4) has this property. This can be seen if one recalls that
Eq.(3.4) is similar to the Schr\"{o}dinger equation of linear harmonic
oscillator such that $(1\slash 8)R_S^2$ takes the place of energy. The
``classically allowed region'' is the one in which
$$
{1\over 8}R_S^2 \geq {1\over 2}\xi^2,\eqno(3.9)
$$
or,
$$
-{1\over 2}R_S\leq \xi\leq{1\over 2}R_S.\eqno(3.10)
$$
Now, if the function $u_n(\xi)$ went to zero in the point
$\xi=-(1\slash 2)R_S$, then that would mean that the wave function of
the harmonic oscillator would be precisely zero in the boundary of its
classically allowed region, which we know is not true. Because the
only physically acceptable solutions to the eigenvalue equation (3.5) are 
those in Eq.(3.6), we
come into the remarkable conclusion that Eq.(3.5) has no solutions
satisfying the given boundary conditions, and we must ask ourselves:
Where is the mistake?

     The mistake is in the boundary condition. If we dismiss the
requirement that $u(a)$ must go to zero when $a$ goes to zero, and
instead state that $u(a)$ goes to zero at some point $a_{min} > 0$,
which is not quite zero, and that $u(a)$ is identically zero for all
$0\leq a< a_{min}$, then -if $a_{min}$ is chosen appropriately- the
functions $u_n(\xi)$ of Eq.(3.6) are solutions to the eigenvalue equation
(3.5), and everything is well.

     As an example, consider the solution, where $n=1$. Because
$H_1(x)=2x$, we have
$$
u_1(\xi)=N_1\xi e^{-{1\over 2}k^2\xi^2}.\eqno(3.11)
$$ 
If we now state the boundary condition in such a way that if $a$ goes
to zero then $u_1(a)$ goes to zero, then that would mean that
$u_1(\xi)$ goes to zero when $\xi$ goes to $\xi_1=-\sqrt{3}(1\slash
k)$, and we readily find that $u_1(\xi)$ does not satisfy this boundary
condition. However, if we state the boundary condition in such a way
that $u_1(a)$ goes to zero when $a$ goes to
$$
a_{min}(1) := \sqrt{3}{1\over k},\eqno(3.12)
$$ 
and vanishes identically for all $a< a_{min}(1)$, then that would
mean that $u_1(\xi)$ goes to zero when $\xi$ goes to zero and vanishes
identically when $\xi< 0$. We readily observe that if we define
$u_1(\xi)$ in such a way that it is the function $u_1(\xi)$ of Eq.(3.11), when
$\xi> 0$, and that it vanishes identically otherwise, then this kind
of $u_1(\xi)$ satisfies the given boundary condition, and it is also a
solution to the eigenvalue equation (3.5). In other words, it is the {\it
physical state} of the quantum black hole, which determines the
boundary condition of the wave function.

    The example we had above gives us a clue to the boundary
conditions of the wave function of a general stationary state of the
Schwarzschild black hole. Since the functions $u_n(\xi)$ of Eq.(3.6)
are the only solutions to the eigenvalue equation (3.5), we must state 
our boundary condition in such a way that the functions $u_n(\xi)$ of
Eq.(3.6) satisfy it. In other words, if we pick up some point
$\tilde{\xi}$, and say that $u_n(\xi)\equiv 0$ for all
$\xi\leq\tilde{\xi}$, then the point $\tilde{\xi}$ must be a zero of
$u_n(\xi)$. 
 
    Now, because every function $u_n(\xi)$ has $n$ zeros, we have
plenty of choice. The most natural choice is to choose the smallest of
these zeros. Since the zeros of the function $u_n(\xi)$ are $1\slash
k$ times the zeros of the Hermite polynomial $H_n(x)$, we denote the
smallest zero of $H_n(x)$ by $h_n$, and state the boundary condition
of the function $u_n(\xi)$ in the following way:
$$
u_n(\xi) \equiv 0\,\,\,\,\,\,\,\,\forall \xi\leq {1\over k}h_n.\eqno(3.13)
$$
Using Eqs.(3.1), (3.4), (3.6) and (3.7) we can at last write the
wave functions $\psi_n(a)$ corresponding to the stationary states of
the Schwarzschild black hole: If
$$
a>{1\over k}(h_n + \sqrt{2n+1}), \eqno(3.14.a)
$$
then
$$
\psi_n(a) = N_n H_n(ka-\sqrt{2n+1})\,{1\over a}\exp\lbrack-{1\over 2}k^2(a-{1\over
k}\sqrt{2n+1})^2\rbrack,\eqno(3.14.b)
$$
and if 
$$
a\leq{1\over k}(h_n + \sqrt{2n+1}),\eqno(3.15.a)
$$
then
$$
\psi_n(a)\equiv 0.\eqno(3.15.b)
$$
Because $H_0$ has no zeros, we must have $n=1,2,3...$
  
    It should be noted that, if our Schr\"{o}dinger equation is true, we
have now solved
the singularity problem of the Schwarzschild black hole. {\it There is no
singularity inside the black hole}. This can be seen from Eq.(3.15),
which states that below some positive value of $a$, which depends on
the quantum state of the black hole, the wave function vanishes
identically. This means that in a given quantum state the
radius $a$ of the throat of the Schwarzschild black hole can never take
values below a certain fixed positive value of $a$. In other words, an
observer inside the Schwarzschild black hole can never fall into the
singularity, where $a=0$, because no such singularity does exist. 

      After this lengthy talk about the boundary conditions of the
wave function we can now go into the physical predictions given by our
model. It follows from Eqs.(3.3) and (3.7) that the eigenvalues of
the area 
$$
A_S = 4\pi R_S^2\eqno(3.16)
$$ 
of the event horizon of the Schwarzschild black hole are:
$$
A_S(n) = (n+{1\over 2}) {{32\pi}\over{c^3}}\hbar G, \eqno(3.17)
$$
where $n=1,2,3...$ When Eq.(3.17) is put in numbers, we get:
$$
A_S(n) = (n+{1\over 2})\times 2.63\times 10^{-68}m^2.\eqno(3.18)
$$
In other words, the area of the event horizon of the black hole can
take only discrete values such that the quanta of the area are of the 
same order of magnitude as the Planck area. This result is similar to
the one given by
Bekenstein and others.$\lbrack 4-8\rbrack$ It is also in harmony with the
results on the quantization of area in quantum gravity in general obtained by
Ashtekar, Rovelli and Smolin.$\lbrack 2\rbrack$
In this paper we obtained the similar result for the Schwarzschild
black hole by means of an explicit calculation.

     From Eqs.(1.3) and (3.17) it follows that the mass eigenvalues of the
Schwarzschild black hole are:
$$
M_n = \sqrt{2n+1}\sqrt{{\hbar c}\over{G}} = \sqrt{2n+1}\times
2.18\times 10^{-8}kg,\eqno(3.19)
$$
and so the corresponding energy eigenvalues are:
$$
E_n = \sqrt{2n+1}\sqrt{{\hbar c^5}\over{G}} = \sqrt{2n+1}\times
1.96\times 10^9 J.\eqno(3.20)
$$
As one can see, the Schwarzschild black hole has a certain ground
state, where $n=1$. The energy of this ground state is
$$
E_1 = 3.39\times 10^9 J.\eqno(3.21)
$$
According to our model, this is the lowest energy state the
Schwarzschild black hole can have.

    It was shown by Hawking long ago that black holes are not
completely black but they can emit radiation.$\lbrack 1\rbrack$ When they emit
radiation they lose their mass. It is now very interesting to see,
what kind of predictions could be made of the spectrum on this so
called {\it Hawking radiation}, using our simple model.  It follows from
Eq.(1.3) 
that when the area is changed from
$A_S$ to $A_S+dA_S$, the mass $M$ is changed from $M$ to $M+dM$ such
that:
$$
dM={{c^4}\over{32\pi G^2}}{1\over M}\,dA_S.\eqno(3.22)
$$
Because it follows from Eq.(3.16) that $dA_S$ is some integer times a
certain area, we see that the energy emitted by a macroscopic black
hole when it performs a transition from state $n_1$ to the state
$n_2=n_1-n$, is:
$$
\Delta E_n = n{{c^3\hbar}\over{G}}{1\over M} = n\times 42.6 Jkg {1\over
M}.\eqno(3.23) 
$$
As an example, let us assume that the mass $M$ of the black hole is
ten solar masses, or $2.0\times 10^{31}kg$. In that case we find that
the energies of the quanta of the Hawking radiation are of the form
$$
\Delta E_n = n\times 1.3\times 10^{-11}eV, \eqno(3.24)
$$
and the corresponding frequencies are:
$$
\nu_n = n\times 3.2kHz.\eqno(3.25)
$$
The curious fact that the differences between the frequencies of the
quanta are of the same order of magnitude as the resolving power of an
ordinary radio receiver, raises some hopes about a possibility to
observe, in some very distant future, genuine quantum gravitational
effects by measuring the spectrum of Hawking radiation, although
this problem, of course, deserves a much more detailed
analysis.$\lbrack 12\rbrack$
    
\bigskip

\bigskip

\bigskip

\centerline{\title 4. Conclusion}

\bigskip

\bigskip

        In our analysis on the quantum mechanical properties of the
Schwarzschild black hole there were three main points. They were an
adoption of the point of view of an observer very far away and at rest
relative to the black hole, the use of the Hamiltonian dynamics of
asymptotically flat spacetimes in a form developed long ago by Regge
and Teitelboim[9], and our decision to describe the gravitational
degrees of freedom of the Schwarzschild black hole by one free
variable. This variable was the radius $a$ of the throat of the black
hole as such as it is observed by an observer in a radial free fall
inside the black hole horizon. By using the formalism of Regge and
Teitelboim, we wrote the classical Hamiltonian $H$ of the
Schwarzschild black hole in terms of $a$ and its canonical momentum
$p$. We then constructed the quantum mechanical Hamiltonian operator $\hat{H}$
corresponding to our minisuperspace-type model from the classical
Hamiltonian $H$. The eigenvalues of $\hat{H}$ are the ADM energies of
the black hole. By writing the eigenvalue equation we obtained the
time-independent Schr\"{o}dinger equation of the black hole.

    Our Schr\"{o}dinger equation, when written in terms of $a$, turned out
to be surprisingly simple. Indeed, we were able to solve that equation
explicitly. The solutions, which gave the mass and energy eigenstates
of the black hole, implied the quantization of the area of the event
horizon in a manner which is in harmony with the results obtained by
Bekenstein and others [4-8]. The result is also entirely in harmony with the
general results of Ashtekar, Rovelli and Smolin on the area
quantization in quantum gravity[2]. Moreover, it was found that the
black hole has a certain ground state in which its mass is non-zero.

      Perhaps the most striking result of our analysis, however, was
the conclusion that there is no singularity inside the black hole
horizon. Indeed, we found that if there were a singularity, then our
Schr\"{o}dinger equation would not have any physically acceptable
solutions. It was also very interesting to observe that, in
a certain sense, the physical states of the quantum black hole
determine their own boundary conditions.

    The physical reliabilty of the results mentioned above depends, of course,
on whether one accepts that the geometrical degrees of freedom of the
Schwarzschild black hole can be described, at least qualitatively, by
means of the variable we used in our analysis, and on whether one
accepts the analysis which lead to the Schr\"{o}dinger equation (2.40).
However, if one accepts the Schr\"{o}dinger equation, one is also
compelled to accept the results, and the fact that at least the
results related to area quantization have also been obtained by
some others, suggests that perhaps our model is not completely erroneous.   

\bigskip

\bigskip

\bigskip

     \centerline{\title Acknowledgment}

\bigskip

\bigskip

      This research was supported by Suomen Kulttuurirahasto, Finland.

\medskip

\medskip

\medskip

\centerline{\title References}

\medskip

\medskip

\medskip

$\lbrack 1\rbrack$ S. W. Hawking, Commun. Math. Phys. {\bf 43} (1975)
199

\medskip

$\lbrack 2\rbrack$ A. Ashtekar, C. Rovelli and L. Smolin, Phys. Rev
Lett. {\bf 69} (1992) 234

\medskip

$\lbrack 3\rbrack$ On the problem of time see, for example, C. J.
Isham, report gr-qc 9210011

\medskip

$\lbrack 4\rbrack$ J. D. Bekenstein, Lett. Nuovo Cimento {\bf 11}
(1974) 467

\medskip

$\lbrack 5\rbrack$ V. Mukhanov, JETP Letters {\bf 44} (1986) 63

\medskip

$\lbrack 6\rbrack$ Yu. I. Kogan, JETP Letters {\bf 44} (1986) 267 

\medskip

$\lbrack 7\rbrack$ M. Maggiore, Nucl. Phys. {\bf B429} (1994) 205

\medskip

$\lbrack 8\rbrack$ U. H. Danielsson and M. Schiffer, Phys. Rev. {\bf
D48} (1993) 4779

\medskip
 
$\lbrack 9\rbrack$ T. Regge and C. Teitelboim, Ann. of Phys. {\bf 88}
(1974) 286

\medskip

$\lbrack 10\rbrack$ See, for example, C. W. Misner, K.Thorne and
J. A. Wheeler, {\it Gravitation} (Freeman, San Francisco, 1973)

\medskip

$\lbrack 11\rbrack$ See, for example, A. Messiah, {\it Quantum
Mechanics}, Vol. 1 (North Holland, Amsterdam, 1961)

\medskip

$\lbrack 12\rbrack$ For a more detailed account on the spectrum of
Hawking radiation, see J. D. Bekenstein and V. F. Mukhanov, Phys.
Lett. {\bf B360} (1995) 7

\bye